\documentclass[english,aps,pra,superscriptaddress,address,showpacs,showacknowledgments]{revtex4}
\usepackage[T1]{fontenc}
\usepackage[latin9]{inputenc}
\usepackage{babel}
\usepackage{amsmath}
\usepackage{amssymb}
\usepackage{epstopdf}
\usepackage{graphicx}
\usepackage{float}
\usepackage{esint}
\inputencoding{utf8}
\usepackage[unicode=true,
 bookmarks=true,bookmarksnumbered=false,bookmarksopen=false,
 breaklinks=false,pdfborder={0 0 1},backref=false,colorlinks=false]
 {hyperref}
\hypersetup{pdftitle={Enhancement of entanglement in distant micromechanical mirrors using parametric interactions}}
\inputencoding{latin9}

\makeatletter
\@ifundefined{textcolor}{}
{%
 \definecolor{BLACK}{gray}{0}
 \definecolor{WHITE}{gray}{1}
 \definecolor{RED}{rgb}{1,0,0}
 \definecolor{GREEN}{rgb}{0,1,0}
 \definecolor{BLUE}{rgb}{0,0,1}
 \definecolor{CYAN}{cmyk}{1,0,0,0}
 \definecolor{MAGENTA}{cmyk}{0,1,0,0}
 \definecolor{YELLOW}{cmyk}{0,0,1,0}
}

\usepackage{babel}
\usepackage{babel}
\usepackage{babel}
\usepackage{babel}
\usepackage{bbold}

\makeatother

\begin{document}

\title{Enhancement of entanglement in distant micromechanical mirrors using
parametric interactions}

\author{Chang-Sheng Hu}
\author{Xi-Rong Huang}
\author{Li-Tuo Shen}
\author{Zhen-Biao Yang}
\author{Huai-Zhi Wu}

\affiliation{Department of Physics, Fuzhou University, Fuzhou 350002, People\textquoteright s Republic of China}

\begin{abstract}
We theoretically investigate the stability of a two cascaded cavity
optomechanical system with optical parametric amplifiers (OPAs) inside
the two coupled cavities, and study the steady-state entanglement
between two distant mechanical resonators. We show that the parameter
regime where the system is unstable without OPAs, such as relatively
high laser intensity and blue detuning, can be exploited
to build the steady-state mechanical entanglement by modulating the
parametric gain. The application of OPAs is helpful to preserve the
mechanical entanglement suffered from the dissipation at some finite
temperature. The scheme provides an alternative way for improving
and engineering the quantum entanglement of two distant mechanical
oscillators.
\end{abstract}
\maketitle

\section{\emph{Introduction}}

Quantum entanglement is a key resource for quantum information processing
and quantum communication. One now has a fairly good understanding
of how to produce entanglement among microscopic systems, such as
atoms \cite{nature_Mandel2003}, ions \cite{nature_Haffner2005}, or artificial qubits \cite{Nature_Barends2014}. In recent
years there has been considerable interest in studying macroscopic
entanglement in the context of cavity optomechanics \cite{arxiv_F Marquardt2009}, where
the great experimental advance has made it possible to study the quantum
effects and the preparation of nonclassical states for the optomechanical
systems, including the realization of the squeezed light \cite{pra_CFabre1994,prx_TPPurdy2013,
njp_AKronwald2014,pra_KenanQu2015, JO_DainiusKilda2015}, squeezed mechanical resonator \cite{prl_GSAngarwal1991,prl_AMari2009,oe_GUWenJu2013, pra_LiaoJieQiao2011,pra_Kronwald2013,pra_Huangsumei2010,pra_LvXinYou2015,
njp_ClerkAA2008, prl_AlexSzorkovszk2013}, macroscopic superposition state \cite{pra_SouBose1999,prl_WilMarshall2003}, and mechanical
entanglement \cite{pra_MPaterno2007, prl_MHartmann2008,njp_Huangsu2009,pra_ZhouLing2011,pra_YinZhangQi2009,prl_WangYingDan2013,pra_WangYD2015, pra_DengZhiJiao2016}.

Macroscopic entanglement of mechanical resonators plays a key role
in testing the fundamental principles of quantum mechanics, quantum
information processing, and ultrahigh-precision measurements \cite{pra_Hofer2011,
prl_Mancini2003,nanoL_YangYT2006,science_Kippenberg2008}. The generation of stationary entanglement with two
or several mechanical objects in an optical cavity has been widely
discussed \cite{pra_Ge2013,pra_Yangchunjie2015,pra_Xuxuewei2013}. Very recently, the coupling scheme between
two cascaded optomechanical setups has also been proposed. The quantum
entanglement between two macroscopic mechanical resonators can be
generated with the systems being connected by an optical fiber or
direct photon tunneling \cite{pra_Joshi2012} , but the mechanical entanglement
is small and limited to the enviromental bath of low temperature. Chen et al.
have found that the entanglement can be enhanced by using a periodically
modulated pump laser \cite{pra_ChenRongXin2014}, where the chirped frequency needs
to be subtlely controlled and is therefore sensitive to fluctuation
of experimental parameters. Further increasing the mechanical entanglement
may require the strong single-photon coupling, and the
mechanical resonators should be firstly cooled down to the quantum
ground state \cite{pra_LiaoJieQiao2014}.

Recent studies have found that including an optical parametric amplifier
inside a cavity could affect the normal-mode splitting behavior between
the coupled movable mirror and the cavity field \cite{pra_Huangsumei2009} and significantly
enhance the optomechanical effective coupling \cite{prl_LvXinYou2015}. Moreover,
squeezing of cavity modes can considerably improve the cooling of
the micromechanical mirror \cite{pra_Huangsumei79_2009} and the squeezing properties can
be transferred from photons to phonons with high efficiency \cite{pra_Agarwal2016}.
Here, we propose a scheme to enhance the mechanical stationary entanglement
for two distant movable mirrors by placing two OPAs inside two coupled
optomechanical cavities. The cavity fields are coupled through a photon-hopping
interaction. It is shown that the limits for the stability condition
for the coupled optomechanical system can be relaxed by controlling
the parametric gain, and therefore, the stable region of the corresponding
parameter space expands. The enhancement of mechanical entanglement
can be achieved by appropriately tuning the parametric gain and the
effective laser-cavity detuning, and is even more pronounced with
laser blue-detuning from cavity resonance. In addition, since the
difficulty of preserving quantum entanglement of macroscopic systems
is often attributed to environment-induced decoherence, we also show
that the OPAs inside the cavities can help to witness the steady-state
entanglement between the mechanical resonators suffered from certain
thermal noise.

The paper is organized as follows. In section II , we introduce the
setup with two coupled cavity optomechanical systems and build the
corresponding theoretical model. In section III, we give the covariance
matrix for the system and use it to calculate the logarithmic negativity,
which quantifies the steady-state mechanical entanglement. In section
IV, we present the effect of the OPAs on the generated mechanical
entanglement by numerical simulation. Finally, we summarize the results
in section V.

\section{MODEL}

We consider a physical setup consisting of two identical Fabry-Perot
cavities, and each cavity contains a degenerate optical parametric
amplifier(OPA) and has one movable and one fixed mirror, as plotted
in Fig. 1. The fixed mirror is partially transmitting, but the movable
mirror is totally reflecting. We assumed that each cavity only has
one resonant mode, and the cavity modes are coupled via a photon-hopping
interaction. The single mode assumption is valid when the
cavity coupling $\lambda$ is small compared with the free spectral
range (FSR) of each uncoupled cavity. The cavity modes with the same
frequency $\omega_{c}$ are driven equally by external lasers with
frequency $\omega_{l}$ and amplitude E . The movable mirrors are
considered as quantum mechanical harmonic oscillators, with the same
effective mass m , frequency $\omega_{m}$ , and damping rate $\gamma_{m}$.
Under the action of cavity-photon-induced radiation pressure, the
movable mirrors will make oscillations around their equilibrium positions.
For the degenerate OPA, we assume that the pump fields interact with
the second-order nonlinear optical crystals at frequency $2\omega_{l}$
, with the gain of the OPAs being $\Lambda$, which depends on the
pumping power. The phase of the pump field is assumed to be $\theta$.
In the rotating frame at the laser frequency $\omega_{l}$, the Hamiltonian
of the system reads $\left(\hbar=1\right)$

\begin{figure}[hbt]
\includegraphics[width=0.7\columnwidth]{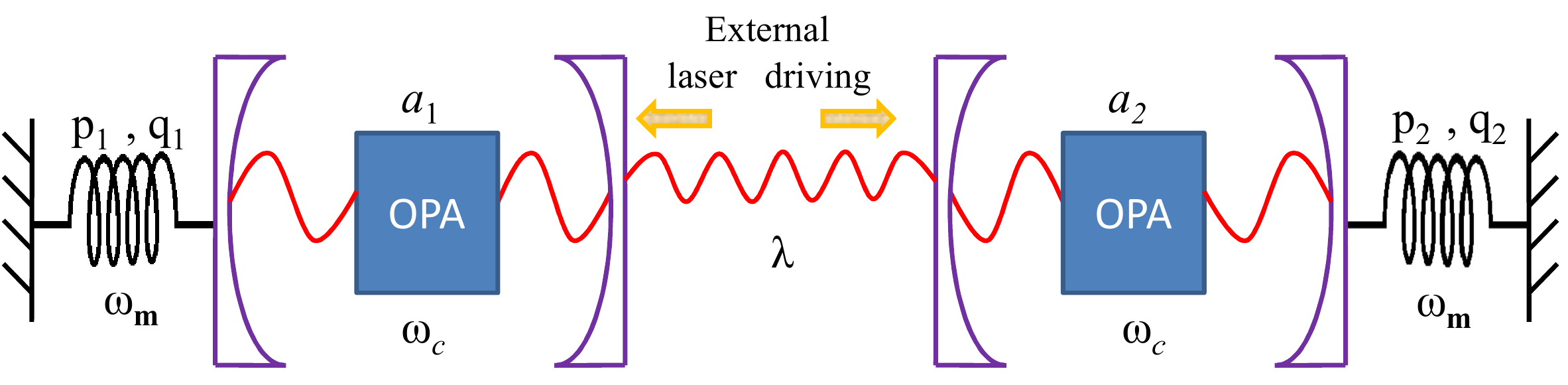}

\caption{\label{fig1}(Color online) Schematic diagram of the cascaded cavity
optomechanical system. Each cavity contains a nonlinear crystal that
is pumped by an external laser. The two optomechanical subsystems
couple to each other via a photon-hopping interaction.}
\end{figure}

\begin{equation}
\begin{array}{cll}
H & = & {\displaystyle \sum_{j=1}^{2}}[\Delta_{0}{\textstyle a_{j}^{\dagger}}a_{j}+\frac{\omega_{m}}{2}{\textstyle (p_{j}^{2}+{\textstyle q_{j}^{2}}})-g{\textstyle a_{j}^{\dagger}}a_{j}q_{j}+iE\text{(}{\textstyle a_{j}^{\dagger}-a_{j})}+i\Lambda({\textstyle a}_{j}^{\dagger2}e^{-i\theta}-{\textstyle a_{j}}^{2}e^{i\theta})]+\lambda({\textstyle a_{1}^{\dagger}a_{2}+a_{1}{\textstyle a_{2}^{\dagger}})},\end{array}
\end{equation}
where  $\Delta_{0}=\omega_{c}-\omega_{l}$ is the laser detuning from
the cavity resonance frequency $\omega_{c}$ ; $a_{j}$ and ${\scriptstyle {\textstyle a_{j}^{\dagger}}}$
are the annihilation and creation operator of the cavity modes, and
$q_{j}$ and $p_{j}$ are the dimensionless position and momentum operators
for the movable mirrors, satisfying the standard canonical commutation
relation $[q_{j},\:p_{j}]=i$. The parameter $g=\omega_{c}x_{ZPF}/L$
is the single-photon coupling strength between the cavity and the
mirror due to radiation pressure, where $x_{ZPF}=\sqrt{\hbar/(2m\omega_{m})}$
is the zero-point fluctuation of the mirror and L is the rest length
of the cavity. $\left|E\right|=\sqrt{2\kappa P/\hbar\omega_{l}}$
is related to the input laser power P , with  $\kappa$ being the cavity
decay rate due to the leakage of photons through the partially transmitting
mirror.

A proper analysis of the system must take the cavity decay and the
mechanical damping into account, according to the Heisenberg equation
of motion , the dissipative dynamics of the full system is described by the following set of nonlinear
quantum Langevin equations (QLEs) \cite{book_Gardiner2004}:

\begin{equation}
\begin{array}{cll}
\dot{q}_{1} & = & \omega_{m}p_{1},\\
\dot{p}_{1} & = & -\omega_{m}q_{1}-\gamma_{m}p_{1}+g{\textstyle a_{1}^{\dagger}a_{1}}+\xi_{1}(t),\\
\dot{q}_{2} & = & \omega_{m}p_{2},\\
\dot{p}_{2} & = & -\omega_{m}q_{2}-\gamma_{m}p_{2}+g{\textstyle a_{2}^{\dagger}a_{2}}+\xi_{2}(t),\\
\dot{a}_{1} & = & -(\kappa+i\Delta_{0})a_{1}+iga_{1}q_{1}+E+2\Lambda e^{i\theta}{\textstyle a_{1}^{\dagger}-i\lambda a_{2}+\sqrt{2\kappa}{\textstyle a_{1}^{in}(t),}}\\
\dot{a}_{2} & = & -(\kappa+i\Delta_{0})a_{2}+iga_{2}q_{2}+E+2\Lambda e^{i\theta}{\textstyle a_{2}^{\dagger}}-i\lambda a_{1}+\sqrt{2\kappa}{\textstyle a_{2}^{in}(t),}
\end{array}\label{eq:QLEs_initial}
\end{equation}
here ${\textstyle a_{j}^{in}}$ are the independent input quantum
noise operators with zero mean value, and their nonzero correlation
function are given by \cite{book_Gardiner2004}

\begin{equation}
\begin{array}[b]{cll}
\left\langle {\textstyle a_{j}^{in\dagger}(t)a_{j}^{in}(t')}\right\rangle  & = & n_{a}\delta(t-t'),\\
{\textstyle \left\langle a_{j}^{in}(t)a_{j}^{in\dagger}(t')\right\rangle } & = & (n_{a}+1)\delta(t-t'),
\end{array}
\end{equation}
with $n_{a}=[\exp(\hbar\omega_{c}/k_{B}T)-1]^{-1}$ being the mean
thermal excitation number in the optical mode, and $k_{B}$ being
the Boltzmann constant. The Hermitian Brownian noise operators ${\textstyle \xi_{1}(t)}$
and $\xi_{2}(t)$ have the quantum statistical properties \cite{pra_Giovannetti2001,rmp_Clerk2010}

\begin{equation}
\begin{array}[b]{rll}
\left\langle \xi_{1}(t)\right\rangle  & = & \left\langle \xi_{2}(t)\right\rangle =0,\\
\left\langle {\textstyle \xi_{i}(t)\xi_{j}(t')}\right\rangle  & = & \frac{\gamma_{m}}{\omega_{m}}\int\frac{d\omega}{2\pi}e^{-i\omega(t-t')}\omega[1+coth(\frac{\hbar\omega}{2k_{B}T})]\delta_{ij},
\end{array}
\end{equation}
where i , j $\in1,\,2$ and $\delta_{ij}$ is the Kronecker delta
function. By assuming the mechanical quality factor $Q_{m}=\omega_{m}/\gamma_{m}\gg1$,
the correlation function of the noise $\xi_{j}(t)$ becomes

\begin{equation}
\left\langle \xi_{j}(t)\xi_{j}(t')+\xi_{j}(t')\xi_{j}(t)\right\rangle /2=\gamma_{m}(2n_{m}+1)\delta(t-t'),
\end{equation}
where $n_{m}=[\exp(\hbar\omega_{m}/k_{B}T)-1]^{-1}$ is the mean thermal
excitation number in the movable mirror.

In the presence of strong laser driving, we can always rewrite each
Heisenberg operator as linear sum of the classical mean$\langle A\rangle=A_{j}^{s}(A=q,\,p,\,a)$
and the quantum fluctuation $\delta A_{j}$ with $\left\langle \delta A_{j}\right\rangle =0$
, i.e.

\begin{equation}
A=A_{j}^{s}+\delta A_{j},\;j=1,2.\label{eq:fluctuation Ope}
\end{equation}

By inserting Eq. (\ref{eq:fluctuation Ope}) into Eq. (\ref{eq:QLEs_initial}),
it is easy to find the steady-state momentum $p_{j}^{s}$ and displacement
$q_{j}^{s}$ of the movable mirrors, and the steady-state amplitude
$a_{j}^{s}$ of the cavity fields given by

\begin{equation}
\begin{array}{llcccll}
p_{1}^{s} & = & p_{2}^{s} & = & 0,\\
q_{1}^{s} & = & q_{2}^{s} & = & q_{s} & = & \frac{g\left|a_{s}\right|^{2}}{\omega_{m}},\\
a_{1}^{s} & = & a_{2}^{s} & = & a_{s} & = & \frac{\kappa-i(\Delta+\lambda)+2\Lambda e^{i\theta}}{\kappa^{2}+(\Delta+\lambda)^{2}-4\Lambda^{2}}E
\end{array}\label{eq:operator mean value}
\end{equation}
with
\begin{equation}
\Delta=\Delta_{0}-gq_{s}=\Delta_{0}-\frac{g^{2}\left|a_{s}\right|^{2}}{\omega_{m}}\label{eq:effecive detuning}
\end{equation}
being the effective input-laser detuning from the cavity resonance
including the frequency shifts caused by the optomechanical interaction \cite{pra_Huangsumei79_2009, pra xuxueweiliyong}, and the linearized quantum Langevin equation for the quantum fluctuation operators as following

\begin{equation}
\begin{array}[b]{lll}
\delta\dot{q}_{1} & = & \omega_{m}\delta p_{1},\\
\delta\dot{p}_{1} & = & -\omega_{m}\delta q_{1}-\gamma_{m}\delta p_{1}+g{\textstyle (a_{s}^{*}\delta a_{1}+a_{s}\delta{\textstyle a_{1}^{\dagger}})}+\xi_{1}(t),\\
\delta\dot{q}_{2} & = & \omega_{m}\delta p_{2},\\
\delta\dot{p}_{2} & = & -\omega_{m}\delta q_{2}-\gamma_{m}\delta p_{2}+g{\textstyle (a_{s}^{*}\delta a_{2}+a_{s}\delta{\textstyle a_{2}^{\dagger}})}+\xi_{2}(t),\\
\delta\dot{a}_{1} & = & -(\kappa+i\Delta_{0})\delta a_{1}+iga_{s}\delta q_{1}+igq_{s}\delta a_{1}+2\Lambda e^{i\theta}\delta{\textstyle a_{1}^{\dagger}}-i\lambda\delta a_{2}+\sqrt{2\kappa}{\textstyle a_{1}^{in}(t),}\\
\delta\dot{a}_{2} & = & -(\kappa+i\Delta_{0})\delta a_{2}+iga_{s}\delta q_{2}+igq_{s}\delta a_{2}+2\Lambda e^{i\theta}\delta{\textstyle a_{2}^{\dagger}}-i\lambda\delta a_{1}+\sqrt{2\kappa}{\textstyle a_{2}^{in}(t),}
\end{array}\label{eq:operator quantum}
\end{equation}
where the nonlinear terms $\delta a_{j}^{\dagger}\delta a_{j}$ and $\delta a_{j}\delta q_{j}$
have been safely neglected \cite{prl_VitaliDavid2007}.

\section{ENTANGLEMENT MEASUREMENT OF THE TWO MECHANICAL OSCILLATORS}

Now, we start looking into the measurement of the steady-state entanglement
between the two mechanical modes. By introducing the amplitude and
phase fluctuations $\delta x_{j}=\frac{1}{\sqrt{2}}(\delta a_{j}^{\dagger}+\delta a_{j})$
and $\delta y_{j}=\frac{i}{\sqrt{2}}(\delta a_{j}^{\dagger}-\delta a_{j})$
for the cavity modes, and $x_{j}^{in}(t)=\frac{1}{\sqrt{2}}(a_{j}^{in\dagger}(t)+a_{j}^{in}(t))$
and $y_{j}^{in}(t)=\frac{i}{\sqrt{2}}(\delta a_{j}^{in\dagger}(t)-\delta a_{j}^{in}(t))$
for the input quantum noises, the linearized QLEs for the quantum
fluctuation in Eq. (\ref{eq:operator quantum}) can be rewritten as
the matrix form

\begin{equation}
\frac{dU(t)}{dt}=\mbox{\ensuremath{MU(t)+N(t),}}
\end{equation}
where$\begin{array}{ccl}
U(t) & = & (\delta q_{1},\delta p_{1},\delta q_{2},\delta p_{2},\delta x_{1},\delta y_{1},\delta x_{2},\delta y_{2})^{T}\end{array}$ is the column vector of continuous variables(CV) fluctuations, $N(t)=(0,\xi_{1}(t),0,\xi_{2}(t),\sqrt{2\kappa}\delta x_{1}^{in}(t),\sqrt{2\kappa}\delta y_{1}^{in}(t),\sqrt{2\kappa}\delta x_{2}^{in}(t),\sqrt{2\kappa}\delta y_{2}^{in}(t))^{T}$
is the column vector of the noise sources, and M is the drift matrix
given by

\[
M=\left(\begin{array}{cc}
M_{1} & M_{2}\\
M_{3} & M_{4}
\end{array}\right),
\]
with
\[
M_{1}=\left(\begin{array}{cccc}
0 & \omega_{m} & 0 & 0\\
-\omega_{m} & -\gamma_{m} & 0 & 0\\
0 & 0 & 0 & \omega_{m}\\
0 & 0 & -\omega_{m} & -\gamma_{m}
\end{array}\right),
\]

\[
M_{2}=\left(\begin{array}{cccc}
0 & 0 & 0 & 0\\
G_{x} & G_{y} & 0 & 0\\
0 & 0 & 0 & 0\\
0 & 0 & G_{x} & G_{y}
\end{array}\right),
\]

\[
M_{3}=\left(\begin{array}{cccc}
-G_{y} & 0 & 0 & 0\\
G_{x} & 0 & 0 & 0\\
0 & 0 & -G_{y} & 0\\
0 & 0 & G_{x} & 0
\end{array}\right),
\]

\[
M_{4}=\left(\begin{array}{cccc}
-\kappa+2\Lambda\cos\theta & \Delta+2\Lambda\sin\theta & 0 & \lambda\\
-\Delta+2\Lambda\sin\theta & -\kappa-2\Lambda\cos\theta & -\lambda & 0\\
0 & \lambda & -\kappa+2\Lambda\cos\theta & \Delta+2\Lambda\sin\theta\\
-\lambda & 0 & -\Delta+2\Lambda\sin\theta & -\kappa-2\Lambda\cos\theta
\end{array}\right).
\]

Here, $G_{x}$ and $G_{y}$ are the real and imaginary parts of the
effective coupling $G=\sqrt{2}ga_{s}$ , respectively. The stability
of the system demands that all the eigenvalues of the drift matrix
M has negative real parts by applying the Routh-Hurwitz criterion
\cite{book_Bellman1964,pra_Dejesus1987}. For the later discussion of the mechanical entanglement
in the present work, the stability condition for all situations are
carefully verified.

As the stability condition is fulfilled, the system will finally
evolve into a unique steady state independent of the initial condition.
Since the dynamics of the fluctuations is linearized, and $\xi_{j}(t)$
and $a_{j}^{in}(t)$ are zero-mean quantum Gaussian noises, the steady
state for the quantum fluctuations must be a Gaussian state \cite{rmp_Weedbrook2012},
for which all the quantum correlations can then be characterized by
the $8\times8$ covariance matrix(CM) V with $V_{ij}=\left\langle U_{i}(\infty)U_{j}(\infty)+U_{j}(\infty)U_{i}(\infty)\right\rangle /2$
, where $U^{T}(\infty)=(\delta q_{1}(\infty),\delta p_{1}(\infty),\delta q_{2}(\infty),\delta p_{2}(\infty),\delta x_{1}(\infty),\delta y_{1}(\infty),\delta x_{2}(\infty),\delta y_{2}(\infty))$
denotes the vector of the fluctuation operators at the steady state $(t\rightarrow\infty$) . Under the stable condition, the CM elements
take the form

\begin{equation}
V_{ij}={\scriptstyle {\displaystyle \sum_{k,l}}}\int_{0}^{\infty}dt\int_{0}^{\infty}dt'W_{ik}(t)W_{jl}(t')\Phi_{kl}(t-t'),\label{eq:CM initial}
\end{equation}
where $W=\exp(Mt)$ , and $\Phi_{kl}(t-t')=\left\langle N_{k}(t)N_{l}(t')+N_{l}(t')N_{k}(t)\right\rangle /2$ is
the matrix of the stationary noise correlation functions. For the
mechanical oscillators with large quality factor $Q_{m}=\omega_{m}/\gamma_{m}\gg1$
, the quantum Brownian noise becomes delta-correlated

\begin{equation}
\begin{array}[t]{ccl}
\Phi_{kl}(t-t') & = & D_{kl}\delta(t-t')\\
 & = & diag[0,\gamma_{m}(2n_{m}+1),0,\gamma_{m}(2n_{m}+1),\kappa(2n_{a}+1),\kappa(2n_{a}+1),\kappa(2n_{a}+1),\kappa(2n_{a}+1)]\delta(t-t'),
\end{array}
\end{equation}
then the Eq. (\ref{eq:CM initial}) can be simplified to

\begin{equation}
V=\int_{0}^{\infty}dtW(t)DW^{T}(t).
\end{equation}

Finally, the CM V satisfies the Lyapunov equation
\cite{prl_Mancini2002}

\begin{equation}
MV+VM^{T}=-D.\label{eq:linear equation}
\end{equation}

The mechanical entanglement can be measured by the logarithmic negativity $E_{N}$
\cite{pra_Vidal2002,pra_Adesso2004}. To calculate $E_{N}$ , we firstly numerically solve
the Lyapunov equation above and find the solution for V , and then
exact the reduced $4\times4$ CM $V_{m}$ for the two mechanical modes
from the full $8\times8$ CM V by just keeping the first four rows
and columns. By rewritting CM $V_{m}$ in the following form

\begin{equation}
\begin{array}{ccc}
V_{m} & = & \left(\begin{array}{cc}
A & C\\
C^{T} & B
\end{array}\right)\end{array}
\end{equation}
with A , B and C being the $2\times2$ sub-block matrices of $V_{m}$
, the logarithmic negativity $E_{N}$ is finally given by
\begin{equation}
E_{N}=\max[0,-\ln\eta^{-}],
\end{equation}
with
\begin{equation}
\eta^{-}=\sqrt{\frac{\Sigma-\sqrt{\Sigma^{2}-4\det V_{m}}}{2},}\label{eq:entanglement}
\end{equation}
 where $\Sigma=\det A+\det B-2\det C$ . If the two mechanical objects
are entangled, the value of $\eta^{-}$ must be less than 1/2 , which
is equivalent to Simon's necessary and sufficient entanglement nonpositive
partial transpose criterion for Gaussian states \cite{EPR}.

\section{NUMERICAL RESULTS AND DISCUSSIONS}

\begin{figure}[H]
\centering
\includegraphics[scale=0.55]{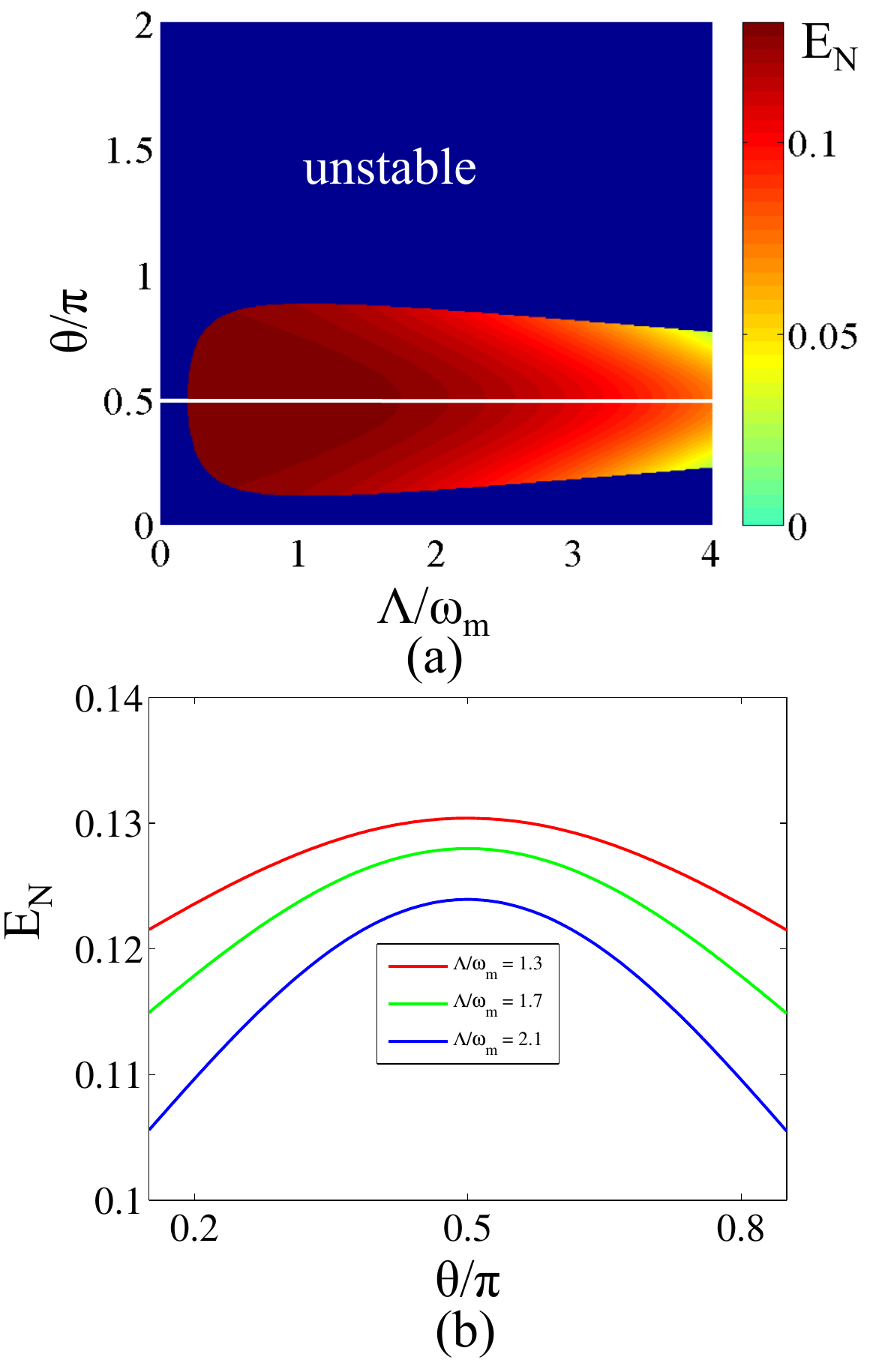}

\caption{\label{fig2}(Color online) (a) The mechanical entanglement $E_{N}$
versus the parametric amplifier phase $\theta$ and parametric gain
$\Lambda$. (b) $E_{N}$ versus the parametric amplifier phase $\theta$
for different parametric gain, $\Lambda/\omega_{m}=1.3$(red curve),
$1.7$(blue curve), $2.1$(black curve). Other parameters are (in
units of $\omega_{m}$): $\kappa/\omega_{m}=0.01$, $\gamma_{m}/\omega_{m}=2\times10^{-3}$,
$n_{a}=0$, $n_{m}=0$, $g/\omega_{m}=4\times10^{-4}$, $\lambda/\omega_{m}=20$,
$E/\omega_{m}=2\times10^{7}$, $\Delta/\omega_{m}=3$ \cite{pra_ChenRongXin2014}.}
\end{figure}

Firstly, we examine the effect of the parametric amplifier phase $\theta$
on the mechanical entanglement $E_{N}$, as shown in Fig. \ref{fig2}.
In all cases, the blue region represents that the system is unstable.
For weak parametric gain $\Lambda$, the mechanical system remains
unstable independent of the amplifier phase. The steady-state mechanical
entanglement can only be achieved for $\Lambda/\omega_{m}>0.2$ and
$\theta$ being limited between $0$ and $\pi$. We can also see from
the Fig. \ref{fig2}(b) that the maximal entanglement $E_{N}$ always
appears at $\theta=\pi/2$ for varied parametric gain ensuring the
stability of the system. This fact is true as well for negative laser
detuning $\Delta$, which is checked numerically. Therefore, in the
following, the parametric amplifier phase will be taken as $\theta=\pi/2$
in all cases.

\begin{figure}[H]
\centering
\includegraphics[scale=0.55]{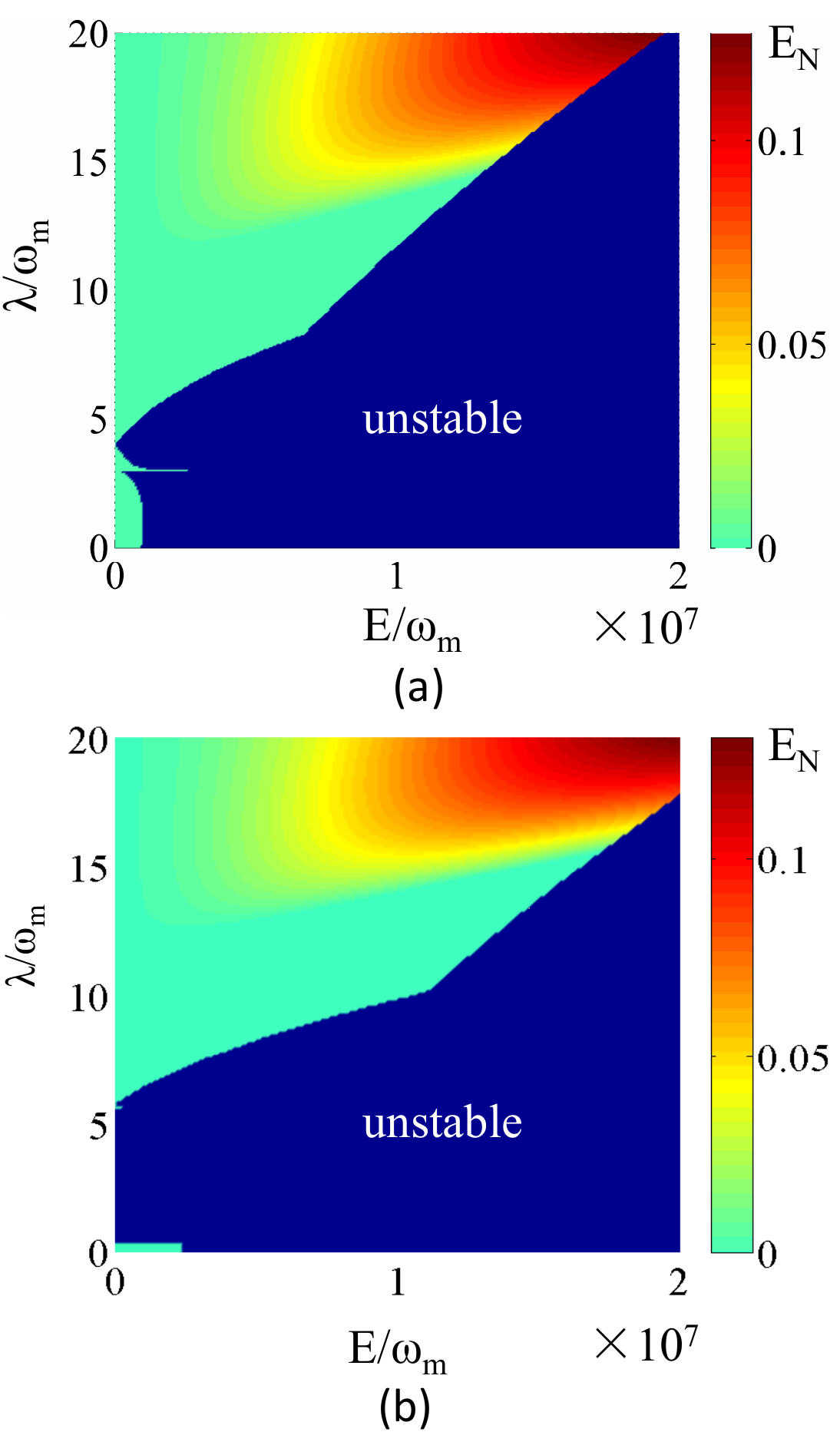}

\caption{\label{fig3}(Color online) The mechanical entanglement $E_{N}$ versus
driving amplitude $E$ and cavity-cavity coupling strength $\lambda$
with the parametric gain (a) $\Lambda/\omega_{m}=0$ and (b) $\Lambda/\omega_{m}=1.3.$
Other parameters are identical to those in Fig. \ref{fig2}. }
\end{figure}

Figs. \ref{fig3}(a) and (b) show the mechanical entanglement $E_{N}$
as a function of the driving amplitude $E$ and the coupling strength
$\lambda$ with OPAs and without OPAs inside the cavities, respectively.
In both cases, it can be seen that the mechanical entanglement $E_{N}$
increases with increasing driving amplitude $E$ and cavity-cavity
coupling strength $\lambda$. For fixed $\lambda$, increasing laser
driving strength does not implies that the mechanical entanglement
will be enhanced. The system becomes unstable when the driving amplitude
$E$ reaches certain limit. However, we find that in the presence
of the OPAs $(\Lambda/\omega_{m}=1.3)$, the stable region for obtaining
the steady-state mechanical entanglement expands in the parameter
space ($\lambda$, $E$), that is the reasonable high steady-state
entanglement can be achieved for smaller cavity-cavity coupling $\lambda/\omega_{m}\sim18$.

\begin{figure}[H]
\centering
\includegraphics[scale=0.5]{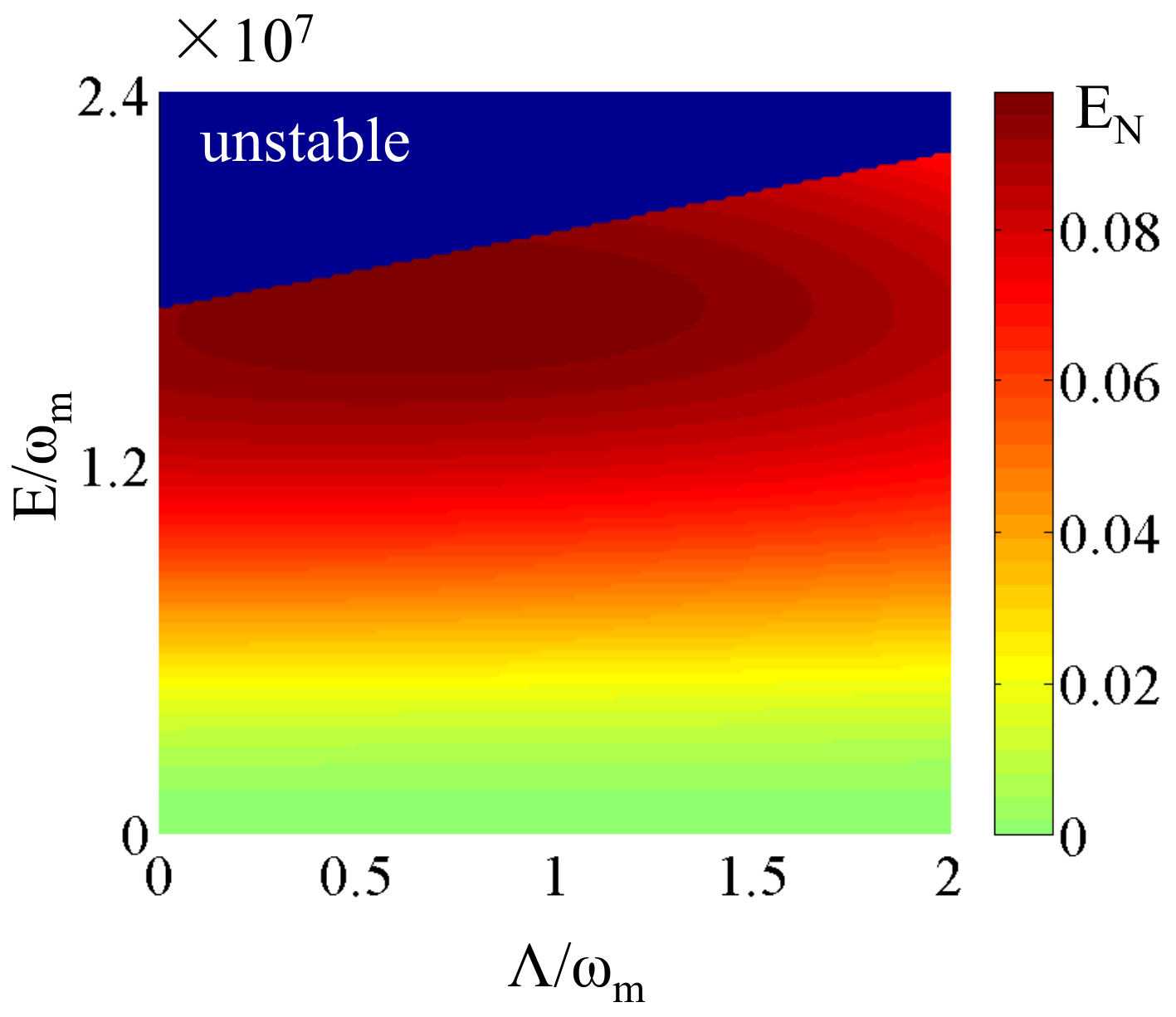}

\caption{\label{fig4}(Color online) The mechanical entanglement $E_{N}$ for
cavity-cavity coupling strength $\lambda/\omega_{m}=18$ versus the
parametric gain $\Lambda$ and the driving amplitude $E$. Other parameters
are same as those in Fig. \ref{fig2}.}
\end{figure}

The limit for the laser driving strength $E$ can be enlarged with
the assistance of OPAs, as can be seen in Fig. \ref{fig4}. While for
parametric gain $\Lambda/\omega_{m}=0$, the optomechanical system
becomes unstable for $E/\omega_{m}\approx1.7\times10^{7}$, the steady-state
entanglement $E_{N}=0.07$ is accessible even for $E/\omega_{m}\approx2.2\times10^{7}$with
parametric gain $\Lambda/\omega_{m}=2$. The maximal entanglement
in this case can be found at $\Lambda/\omega_{m}=0.85$ and $E/\omega_{m}=1.71\times10^{7}$.

\begin{figure}[H]
\centering
\includegraphics[scale=0.5]{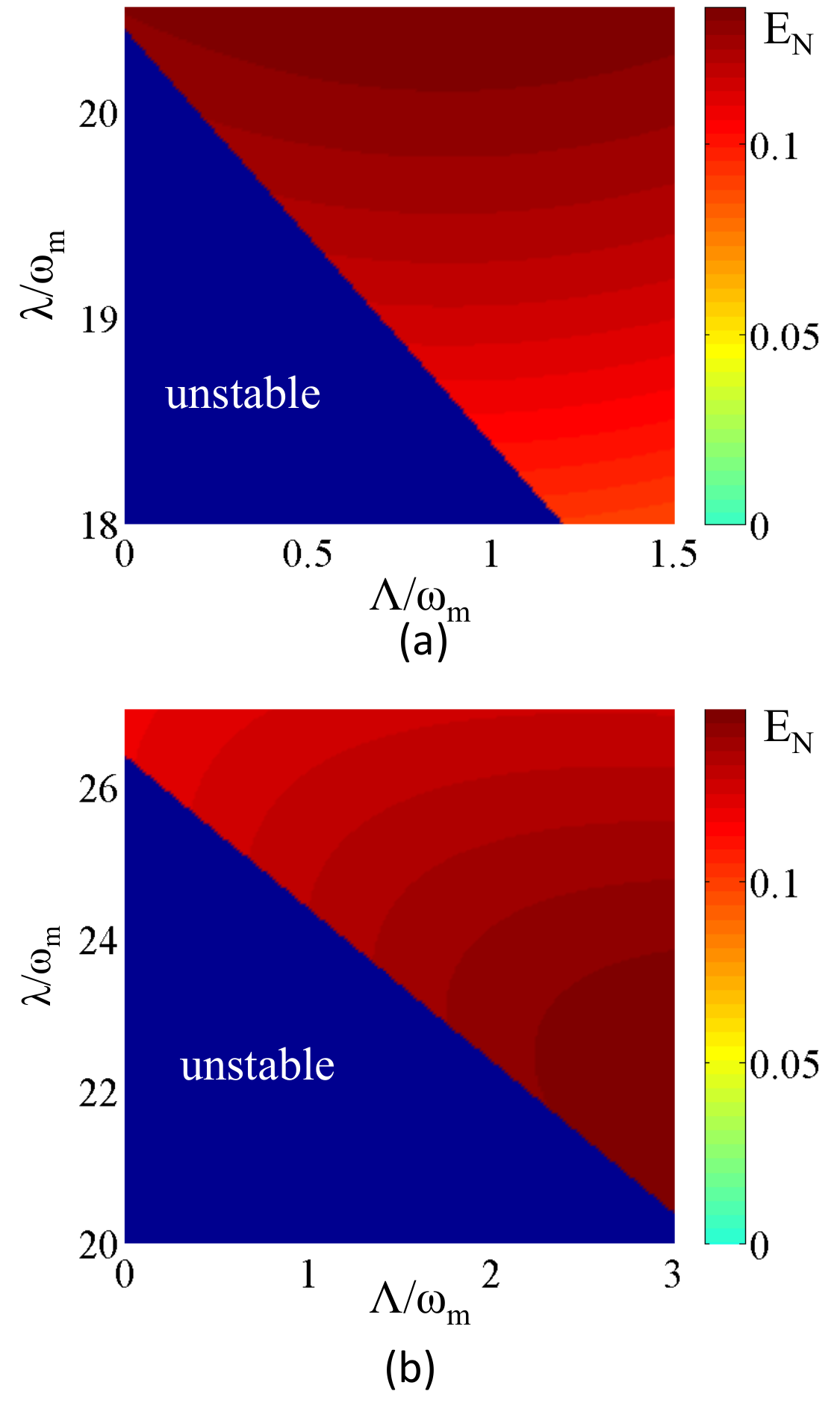}

\caption{\label{fig5}(Color online) The steady-state mechanical entanglement
$E_{N}$ versus parametric gain $\Lambda$ and cavity-cavity coupling
strength $\lambda$, and the effective cavity detuning (a) $\Delta/\omega_{m}=3$
and (b) $\Delta/\omega_{m}=-3$. Other parameters are the same as
in Fig. \ref{fig2}.}
\end{figure}

As we have seen in the Fig. \ref{fig3}, the steady-state mechanical
entanglement is also limited by the cavity-cavity coupling $\lambda$,
namely, the larger the cavity-cavity coupling is, the stronger laser
driving strength required. However, this limit can also be released
by using the OPAs, as shown in Fig. \ref{fig5}, where the steady-state
mechanical entanglement $E_{N}$ for positive laser detuning $\Delta/\omega_{m}=3$
{[}Fig. \ref{fig5}(a){]} and negative laser detuning $\Delta/\omega_{m}=-3$
{[}Fig. \ref{fig5}(b){]} as a function of the parametric gain $\Lambda$
and the coupling strength $\lambda$ are plotted. Setting the driving
amplitude $E/\omega_{m}=2\times10^{7}$, in the absence of OPAs, the
fulfillment of the stability condition is impossible for $\lambda/\omega_{m}<20.4$
with $\Delta/\omega_{m}=3$, and for $\lambda/\omega_{m}<26.4$ with
$\Delta/\omega_{m}=-3$. The involvement of the OPAs with finite parametric
gain can help to stably entangle the two mechanical oscillators under
the condition of less cavity-cavity coupling $\lambda$, which may
be significant for relieving experimental constraints. Moreover, the
steady-state mechanical entanglement $E_{N}$ are enhanced by the
parametric gain in both cases for the same cavity-cavity coupling.

\begin{figure}[H]
\centering
\includegraphics[scale=0.5]{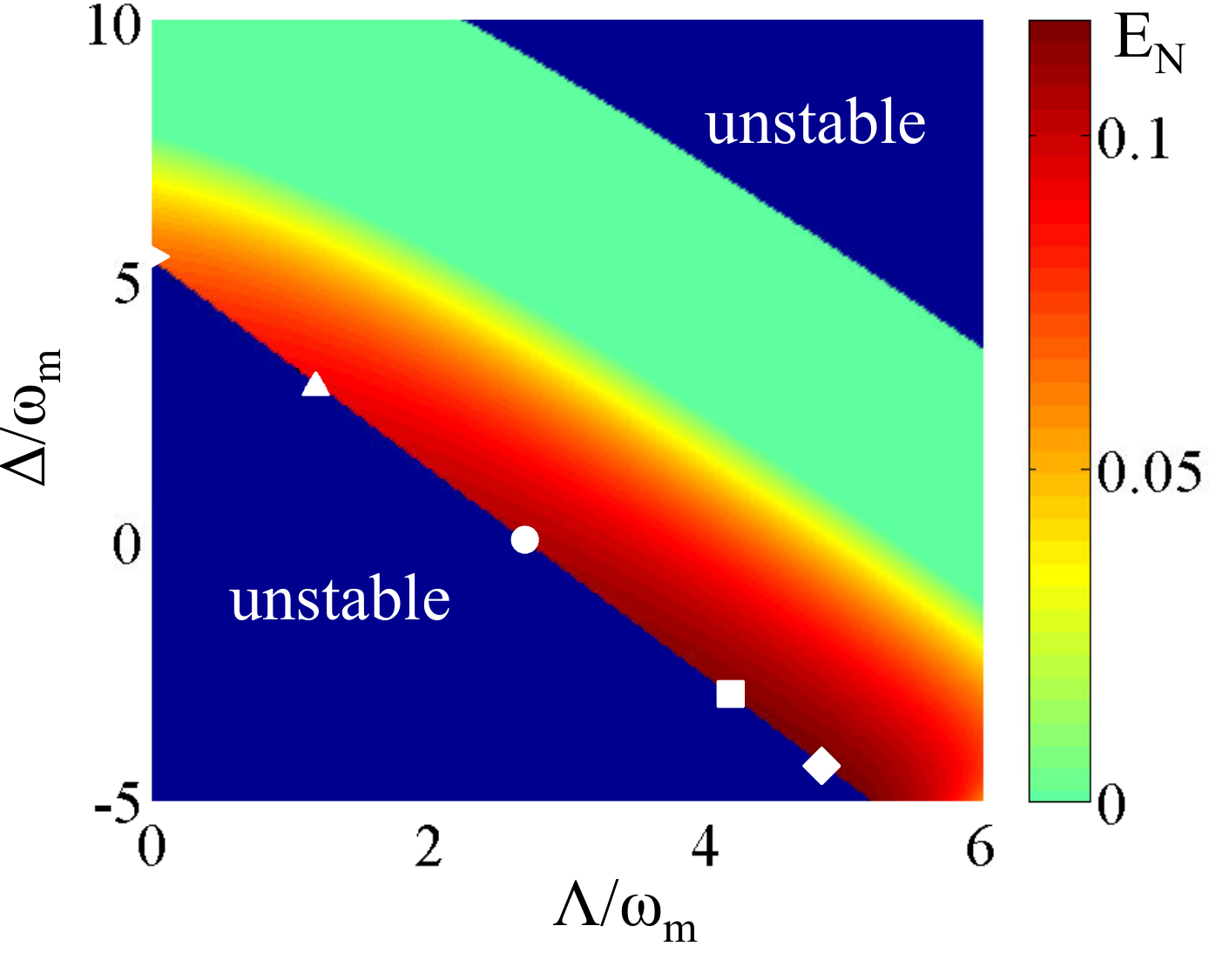}

\caption{\label{fig6}(Color online) The mechanical entanglement $E_{N}$ versus
parametric gain $\Lambda$ and the effective cavity detuning $\Delta$.
Other parameters are same as the Fig. \ref{fig5}. The coordinate
data at $(\Lambda/\omega_{m}=0,\Delta/\omega_{m}=5.48,E_{N}=0.066)$,
$(\Lambda/\omega_{m}=1.18,\Delta/\omega_{m}=3.03,E_{N}=0.091),$
$(\Lambda/\omega_{m}=2.71,\Delta/\omega_{m}=0,E_{N}=0.106),$ $(\Lambda/\omega_{m}=4.19,\Delta/\omega_{m}=-2.99,E_{N}=0.115),$
and $(\Lambda/\omega_{m}=4.83,\Delta/\omega_{m}=-4.25,E_{N}=0.117)$
are indicated, respectively, by right triangle, upper triangle , circular,
square, and diamond.}
\end{figure}

\begin{figure}[H]
\centering
\includegraphics[scale=0.33]{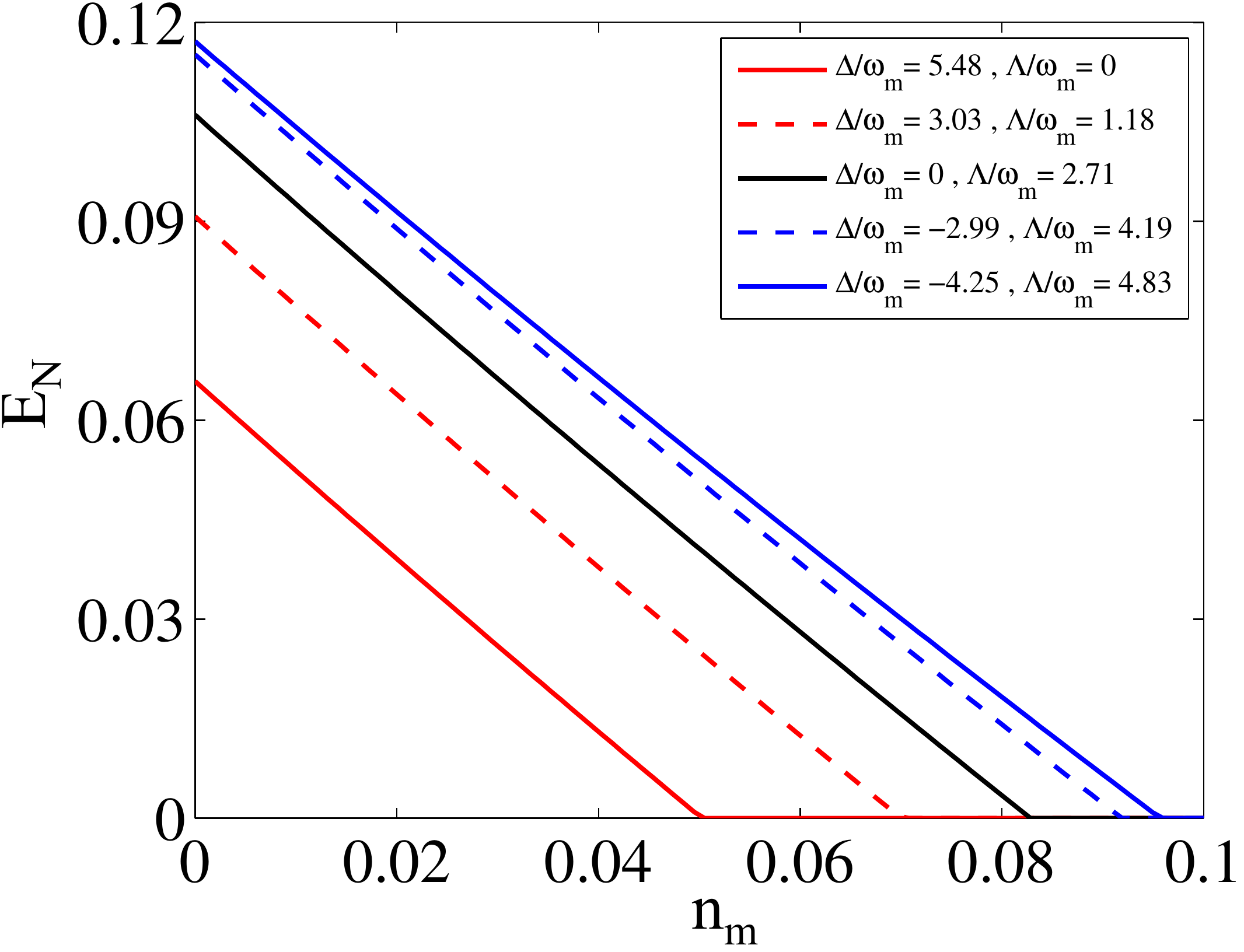}

\caption{\label{fig7}(Color online) The mechanical entanglement $E_{N}$ versus
the thermal phonon number $n_{m}$. The curves are corresponding to
the different coordinates indicated in Fig. \ref{fig5}. Other parameters
are same as the Fig. \ref{fig6}.}
\end{figure}

The effect of the OPAs on the steady-state mechanical entanglement
can be further seen in Fig. \ref{fig6}, where we show the mechanical
entanglement $E_{N}$  versus the parametric gain $\Lambda$ and the
effective cavity detuning $\Delta$. When the OPAs are absent $(\Lambda/\omega_{m}=0)$,
the maximal mechanical entanglement can only reach $E_{N}=0.066$
for the effective detuning around $\Delta/\omega_{m}=5.48$. Then,
it becomes nearly double $E_{N}=0.117$ at $\Delta/\omega_{m}=-4.25$
with the parametric gain $\Lambda/\omega_{m}=4.83$. On the other
hand, the stability of the system is confined in a band area with
the increasing the parametric gain $\Lambda$. For $\Lambda/\omega_{m}>2.71$,
the stability can be found even at negative detuning $\text{\ensuremath{\Delta}}$,
where the input laser may be blue-detuned from the cavity resonance
(i.e. $\Delta_{0}/\omega_{m}<0$ ). This intriguing effect is very
different from the situation without OPAs \cite{pra_Joshi2012}. The
mechanical entanglement $E_{N}$ as a function of the average thermal
occupancy of the two mirrors $n_{m}$ is shown in Fig. \ref{fig7},
where we find that $E_{N}$ almost decreases linearly as $n_{m}$
increases. Thus, for a non-zero thermal reservoir for the mechanical
oscillators, the well selected laser detuning and parametric gain
can help to preserve the mechanical entanglement.

\section{CONCLUSIONS}

In conclusion, we have discussed the stability of the coupled optomechanical
systems and the enhancement of mechanical entanglement for the two
movable mirrors by using the OPAs. We have found that the stable region
in the parameter space consisting of photon-hopping rate and laser
driving strength can be enlarged due to the parametric amplification.
Comparing with the case without OPAs, a well-selected parametric gain
can help to stabilize the system for increasing laser intensity and
negative laser detuning, and enhance the steady-state mechanical entanglement.
The application of OPAs can be good for preserving the mechanical
entanglement suffered from the decoherence induced by certain temperature
limit of the environment. The scheme discussed here provides an alternative
method for improving the quantum entanglement of two distant mechanical
oscillators in addition to quantum reservoir engineering and periodic
modulation with chirped driving laser.

\begin{acknowledgments}
This work is supported by the Major State Basic Research Development
Program of China under Grant No. 2012CB921601; the National Natural
Science Foundation of China under Grants Nos. 11305037, 11347114,
11374054 and 11422437. The fund from FuZhou University CSH thanks
Ze-Lin Zhang and Du Ran for useful discussions.\end{acknowledgments}

\end{document}